\documentclass[twocolumn,showpacs,preprintnumbers,amsmath,amssymb,prl]{revtex4}
\usepackage{graphicx}
\usepackage{dcolumn}
\usepackage{bm}

\begin{document}
\title{Measurement of the $2S$ Hyperfine Interval
in Atomic Hydrogen}

\author{N.\,Kolachevsky}
\altaffiliation[Also at P.N. Lebedev Physical Institute, Moscow,
Russia] \
\author{A.\,Matveev}
\altaffiliation[Also at P.N. Lebedev Physical Institute, Moscow,
Russia] \
\author{J.\,Alnis}
\author{C.G.\,Parthey}
\author{S.G.\,Karshenboim}
\altaffiliation[Also at D.I. Mendeleev Institute for Metrology,
St. Petersburg, Russia] \
\author{T.W. H\"{a}nsch}
\affiliation{Max-Planck-Institut f\"{u}r Quantenoptik,
85748 Garching, Germany}

\date{\today}

\begin{abstract}
An optical measurement of the $2S$ hyperfine interval in atomic
hydrogen using two-photon spectroscopy of the $1S$-$2S$ transition
gives a value of 177\,556\,834.3(6.7)\,Hz. The uncertainty is
 2.4 times smaller than achieved by our group in 2003 and more
than 4 times smaller than for any independent radio-frequency
measurement. The specific combination of the $2S$ and $1S$
hyperfine intervals predicted by QED theory $D_{21}=8 f_{\rm
HFS}({2S}) - f_{\rm HFS}({1S})=48\,953(3)$\,Hz is in good
agreement with the value of $48\,923(54)$\,Hz obtained from this
experiment. 
\end{abstract}

\maketitle

Precision spectroscopy in simple atomic systems and predictions by
quantum-electrodynamics theory (QED) supply essential data for
determination of fundamental constants (\cite{Mohr08, Biraben08,
Hori06}). The measurements also enable sensitive tests of the
theory ensuring high confidence in calculations. One type of these
tests is based on the hyperfine splitting (HFS) measurements in
hydrogen-like systems.

In conventional atoms the sensitivity of the HFS-based tests is
restricted by an insufficient knowledge of their nuclear structure
\cite{Eides01}. The corresponding uncertainty can be reduced by
construction of the specific difference of the $2S$ and $1S$ HFS
frequencies $D_{21} = 8 f_{\rm HFS}({2S}) - f_{\rm HFS}({1S})$ for
which the nuclear size effects significantly cancel out
\cite{Zwanziger61,Sternheim63,Karshenboim05}. The interest in the
$D_{21}$ calculations was inspired by experiments of P.\,Kusch
{\it et al.} \cite{Kusch56} and significant progress has been made
recently \cite{Karshenboim02}. The present theoretical uncertainty
is due to fourth-order QED corrections such as
$\alpha(Z\alpha)^2m/M$ $\alpha^2(Z\alpha^2)$ in units of $f_{\rm
HFS}({1S})$ ($\alpha$ is the fine structure constant, $Z$ is the
nuclear charge, $m/M$ is the electron-to-nucleus mass ratio) which
are related to HFS tests in muonium, spectroscopy in positronium
and the Lamb shift in H \cite{Karshenboim05}.

 The $1S$ and
$2S$ HFS frequencies are accurately measured in  H, D, and
$^3$He$^+$. The lowest $D_{21}$ relative uncertainty of 10\,ppb
(normalized by $f_{\rm HFS}({1S})$) is reached in the He$^+$ ion
\cite{Schluesser69,Prior77}, while for H and D the uncertainty is
about 100\,ppb. The present sensitivity of these QED tests is
mostly restricted by the experimental uncertainty of $f_{\rm
HFS}({2S})$, since $f_{\rm HFS}({1S})$ is measured with much
higher accuracy.

In 2003 we implemented an optical method for measuring the $2S$
HFS frequency in H by two-photon spectroscopy of the $1S$-$2S$
transition \cite{Kolachevsky04}. The result of
$177\,556\,860(16)$\,Hz improved the previous values
\cite{Kusch56,Hessels00}
measured by radio-frequency spectroscopy. The calculation of
$D_{21}$ from the two most recent measurements
\cite{Hessels00,Kolachevsky04} and the precisely measured value of
 $f_\textrm{HFS}(1S)=1420\,405\,751.768(1)$\,Hz
\cite{Ramsey93} showed a deviation from the theoretical prediction
($D_{21}^\textrm{theor}=48.953(3)$\,kHz, \cite{Karshenboim02}) at
the $2\sigma$ level ($\sigma$ is the $D_{21}$ uncertainty).

In this work we have re-measured $f_\textrm{HFS}(2S)$ in H. The
optical method relies on the measurement of the frequency
difference between the two two-photon transitions $1S(F=0)
\rightarrow 2S(F=0)$ (the {\it singlet}, $f_\textrm{s}$) and
$1S(F=1) \rightarrow 2S(F=1)$ (the {\it triplet}, $f_\textrm{t}$)
recorded sequentially in time \cite{Kolachevsky04}. The splitting
is obtained from  $f_{\rm HFS}({2S}) = f_{\rm
HFS}({1S})+f_\textrm{t}-f_\textrm{s}$ in zero magnetic field.

As the most critical improvement compared to \cite{Kolachevsky04}
we use a new ultra-stable optical frequency reference
\cite{Alnis08}. Also, a re-analysis of the 2S HFS frequency
pressure shift shows that it is negligible in our apparatus.

\begin{figure}[t!]
\begin{center}
\includegraphics [width=0.9\columnwidth]{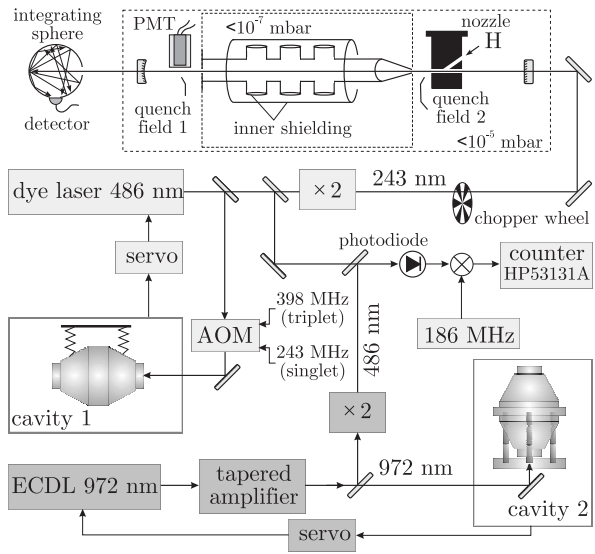}
\caption{Experimental setup. The dye laser locked to the ULE
cavity 1 in horizontal configuration serves for spectroscopy of
the $1S$-$2S$ transition in H. To switch between the {\it singlet}
and the {\it triplet} transitions the double-pass AOM frequency is
changed. The 972\,nm ECDL is continuously locked to the ULE cavity
2 in vertical configuration and serves as an optical frequency
reference. The beat note frequencies of 32\,MHz ({\it triplet})
and 29\,MHz ({\it singlet}) are recorded by the counter
simultaneously with the $1S$-$2S$ data.
 }\label{fig1}
\end{center}
\end{figure}

 To measure $f_\textrm{HFS}(2S)$ we
sequentially excite the {\it singlet} and the {\it triplet}
transitions by the second harmonic of a 486\,nm dye laser
\cite{Fischer04} locked to the Ultra-Low Expansion (ULE) glass
reference cavity 1 with horizontal axis (Fig.\ref{fig1}). The
dye laser has a line width of 60\,Hz (for 0.2\,s) and a frequency
drift of about 1\,Hz/s while its frequency  stability is
$5\times10^{-14}$ for $10^3$\,s (linear drift corrected).

To tune the dye laser frequency between the two hyperfine
transitions,
 a double pass acousto-optic modulator (AOM) is
installed between cavity\,1 and the laser. The required
 frequency detuning of $310$\,MHz is too big to  tune the laser
rapidly without taking it out of lock.

We take advantage of the excellent frequency stability of an
external cavity diode laser (ECDL) at 972\,nm locked to the
thermally- and vibrationally-compensated ULE cavity 2 described in
details in \cite{Alnis08}.  It is stabilized at the zero expansion
temperature such that the influence of external temperature
fluctuations are strongly suppressed. The frequency drift is
nearly linear with a slope of about +50 mHz/s. The ECDL has a line
width of 0.5\,Hz and a frequency stability of $4\times10^{-15}$ in
$10^3$\,s. The  ECDL is continuously locked to cavity 2. Its
frequency is monitored by a fiber frequency comb referenced to an
active H-maser to ensure correct operation.

The beatnote frequency between the dye laser and the second
harmonic of the ECDL is mixed down with a local oscillator
(186\,MHz) to the 30\,MHz range and then counted with a counter as
shown in Fig.\ref{fig1}. All oscillators are referenced to the
GPS-disciplined H-maser guaranteing a fractional frequency
uncertainty  of $10^{-14}$.

The measurement of the beatnote frequency shows that the
re-locking of the dye laser causes two effects: First, after
re-locking, the dye laser frequency shows a strong non-linear
drift up to a few Hz/s due to transient thermal effects in
cavity\,1. Second, it causes abrupt random changes of the laser
frequency of up to 100\,Hz at 486\,nm due to insufficient servo
loop amplification at low frequencies. In 2003 we had no
possibility to monitor the dye laser frequency during the
measurement. The laser instability resulted in excessive data
scatter and could cause a systematic shift. In the new measurement
the instability of the dye laser has basically no influence on the
data quality, as the ECDL is never unlocked.

The second harmonic of the dye laser is coupled to a linear
enhancement cavity forming a standing wave for Doppler-free
two-photon spectroscopy. Atomic hydrogen produced in a microwave
 discharge is cooled down to 4-7\,K by a copper nozzle and escapes
 along the cavity axis. Some of the atoms are excited
to the $2S$ state during their flight. The $2S$ state is quenched
in the detection region at a distance of 19.5\,cm from the nozzle
 by an electric field of 10\,V/cm (quench field
1). The $2S$ population is measured by counting the Lyman-$\alpha$
photons. Compensation coils together with $\mu$-metal shielding
 suppress magnetic fields in the interrogation volume to less than 30\,mG.
A part of the volume (18\,cm in length)  is surrounded by
additional magnetic shielding with a suppression factor of 300
which also serves as beam collimator and Faraday cage. This
shielding separates the high-vacuum region
 from the rest of the vacuum chamber. All parts of the apparatus surrounding the beam are coated with
graphite to suppress stray electric fields.

Compared to \cite{Kolachevsky04} we use a cryogenic nozzle with
bigger diameter of 2.2\,mm which allows recording H lines up to
1\,hour without melting the frozen layer of H$_2$ on its inner
surface. The film improves the $2S$ count rate, but at some
thickness it blocks the 243\,nm laser beam and should be melted.
 An electrode
 is installed close to the nozzle to
optionally quench the $2S$ atoms excited within that region (quench field
2).

We use time-of-flight detection to study different velocity groups
in the cold atomic beam. The 243\,nm light is chopped at 160\,Hz,
and a multi-channel scaler records counts falling into 12 time
bins starting at $\tau=10,\,210\,,..., 2210\,\mu$s and finishing
at 3\,ms after the 243\,nm light is closed by the chopper.

The UV power in the enhancement cavity is maintained nearly
constant at the level of 300\,mW per direction is monitored by a
calibrated photodiode installed after the output coupler. To
prevent beam-pointing effects we use an integrating sphere. The
power fluctuations are taken into account by correction of the
measured frequencies for the AC Stark shift which is evaluated by
a Monte-Carlo code \cite{Kolachevsky06}. For $\tau=810\,\mu$s the
AC Stark shift correction equals 1.3(1)\,Hz/mW.

 The measurement sequence is the same
as in \cite{Kolachevsky04}: Groups of 2-4 {\it singlet} or {\it
triplet} spectra are recorded one after another, the time $t_0$
and the AOM frequency corresponding to each line center are
defined by a fit. We use either a Lorentzian fit or an
unsymmetrical line shape obtained by averaging all superimposed
 and amplitude-normalized experimental line shapes for each $\tau$ (``averaged line fit'').
The lines are fitted by the averaged line fit for the given $\tau$
using 3 parameters:
 the amplitude $A$, the frequency
 offset $f_0$ and the background. The difference $f_\textrm{t}-f_\textrm{s}$ is obtained
 from a double linear fit of 4 neighboring groups of values
 $f_0(t_0)$ after the AC
Stark shift correction.  A constant linear drift of the ECDL
frequency is assumed during
 recording of each 4 groups.

\begin{figure}[t!]
\begin{center}
\includegraphics [width=0.85\columnwidth]{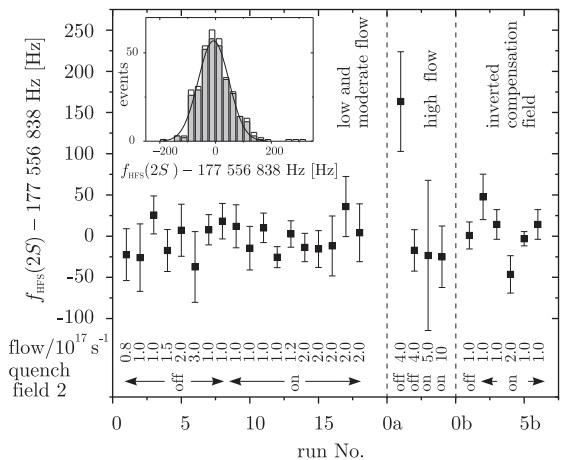}
\caption{$2S$ HFS frequency in H evaluated for $\tau=810\,\mu$s
using the averaged line fit. Each data point is the average over
one set of measurements at unchanged conditions (particle flow,
quench field 2, compensation magnetic field). The histogram shows
$f_\textrm{HFS}(2S)$ data for low and moderate particle flows
(gray bars) and all data (white bars) except the ones taken at
inverted compensation field. The solid line is a Gaussian
fit.}\label{fig2}
\end{center}
\end{figure}

During 17\,days of measurement in February-April 2008 about 1200
$1S$-$2S$ hydrogen spectra have been recorded in 28 sets
(Fig.\,\ref{fig2}). Three types of tests were performed: (i)
variation of the particle flow coming to the cold nozzle in the
range (0.8-10)$\times10^{17}\,\textrm{s}^{-1}$, (ii) the quench
field\,2 switched on/off, (iii) the direction of the compensation
magnetic field is reversed. Further we consider the most important
systematic effects.

\paragraph{The collisional shift.}  In 2003 we
 set the upper bound for the $2S$ HFS interval frequency shift as the
 total shift of the $2S$ level (8\,MHz/mbar,
 \cite{Kolachevsky04}). Now we theoretically re-analyze the shift which appears only
 in third order of perturbation theory if collisions
 with H($1S$) are considered \cite{Dutta70}.
 For each of the excited $P$-states of the
 colliding partners $a$ and $b$ ($m$ and $n$ correspondingly) the
 contribution to the shift of the $2S$ hyperfine component scales inversely to the sum
 $E^{a}_{2S}-E^{a}_{mP}+E^{b}_{1S}-E^{b}_{nP}$. The differential
 HFS
shift is on the order of $10^{-7}$ of the $2S$ level shift due
 to the small ratio of  the $2S$ HFS energy and the difference $E^{b}_{1S}-E^{b}_{nP}$.
  The short-range interaction does not
 contribute since for impact parameters smaller than $20a_0$ ($a_0$ is the Bohr
 radius) the $2S$ state quenches.
 Integration over the discrete spectrum gives a result on the
 order of 10\,Hz/mbar, the continuum gives a similar
 contribution. The effect from collisions with H$_2$ should be of the
same size because the closest dipole transition from the H$_2$
ground state lies is the UV region.
 The intra-beam
pressure in the nozzle is $10^{-4}\,$mbar and rapidly decreases in
the expanding beam so we can neglect the shift.  A stronger effect
may be induced by the dipole interaction with photoionized H
\cite{Prior77}. Since there are only 10 protons present in the
excitation zone at a time the effect is negligible.

\begin{figure}[t!]
\begin{center}
\includegraphics [width=0.8\columnwidth]{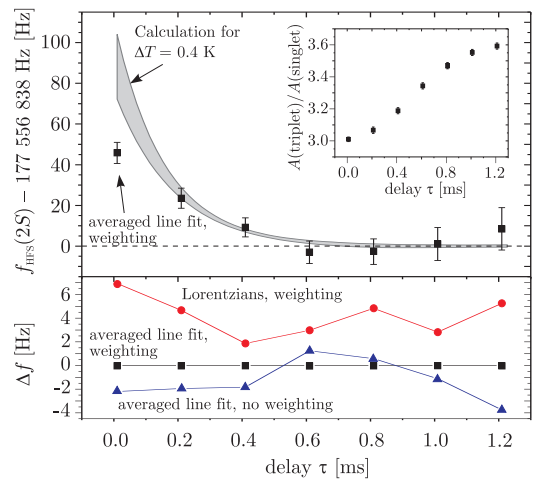}
\caption{Top: Averaged experimental $f_\textrm{HFS}(2S)$ results
for different $\tau$ (squares). Only low and moderate flow data
from Fig.\,\ref{fig2} are analyzed. The gray $1\sigma$ area shows
the expected frequency shift for  $\Delta T=0.4$\,K. Inset: The
measured amplitude ratio of the {\it triplet} and {\it singlet}
transitions which is consistent with $\Delta T=0.4$\,K. Bottom:
mutual deviation of averaged HFS interval frequencies for
different evaluation methods. Error bars are similar to the ones
from the upper plot.}\label{fig3}
\end{center}
\end{figure}

We extrapolate the data of
 Fig.\,\ref{fig2} to zero flow separately for each of the time bins.
 No systematic deviation between the extrapolated and the averaged values is observed. For  $\tau=810\,\mu$s
 the difference equals 1\,Hz with uncertainties of 11\,Hz and 6\,Hz. The data taken
 at higher flows have an excessive scatter due to an instability of the
overloaded cryogenic vacuum system,
 so we exclude
 them (shifting the final value by $+0.5\sigma$) from the analysis. The data are averaged
  without adding any systematics.

\paragraph{Line shape\,/\,beam temperature.} We analyze
the averaged low- and moderate-flow data (Fig.\,\ref{fig2}) for
different delays $\tau$ (Fig.\ref{fig3}, top) which shows an
increase of the measured $f_\textrm{HFS}(2S)$ frequency for
shorter delays.The differential 2nd order Doppler effect is on the
mHz level and cannot explain the difference. The analysis of the
line shapes indicates that the $2S(F=0)$ atoms have a higher
temperature $T$ than the $2S(F=1)$ ones (assuming a Maxwellian
distribution). The amplitude ratio of the two hyperfine
transitions varies from 3.1 to 3.6 depending on $\tau$ (see the
inset) which means that the fraction of  slow $\it singlet$ atoms
is less than for the $\it triplet $ ones.

Fitting the {\it singlet} and {\it triplet} lines by line shapes
simulated for different beam temperatures \cite{Kolachevsky06} we
find for the difference $\Delta T=\langle
T_\textrm{s}\rangle-\langle T_\textrm{t}\rangle=0.4$\,K with
$\langle T_\textrm{s}\rangle=4.2$\,K. The thermalization of
hydrogen atoms on the H$_2$ film depends on its spin state which
we still cannot explain.

We have evaluated the expected shift of $f_\textrm{HFS}(2S)$ {\it
vs.} $\Delta T$ for different $\tau$ by Monte-Carlo simulations.
The dependencies are nearly linear with slopes of $220(40)$\,Hz/K
($10\,\mu$s), $68(5)$\,Hz/K ($210\,\mu$s), $17(4)$\,Hz/K
($410\,\mu$s), $5(3)$\,Hz/K ($610\,\mu$s), and $0(2)$\,Hz/K for
higher delays (Fig.\,\ref{fig3}). The uncertainties result from
numerical errors and possible jitter of $\tau$ at a level of
$20\,\mu$s. As expected, the effect vanishes at higher $\tau$
since the velocity distribution of the delayed atoms is
insensitive to the initial distribution. The optimal compromise
between the statistical uncertainty and the effect of different
beam temperatures is reached at $\tau=810\,\mu$s.

The influence of the fitting model and data weighting is analyzed
in Fig.\,\ref{fig3} (bottom).  The Lorentzian fit of strongly
asymmetrical spectra for $\tau=10\,\mu$s results in a shift of
7(10)\,Hz. On the other hand, the line shape for $\tau=810\,\mu$s
is indistinguishable from a Lorentzian, so the effect reduces to
the sub-hertz level. For $\tau=810\,\mu$s we get
$f_\textrm{HFS}(2S)$ of $177\,556\,835.3(6.2)$\,Hz and add 2\,Hz
uncertainty for line shape\,/\,beam temperature shifts.

\begin{figure}[t!]
\begin{center}
\includegraphics [width=0.75\columnwidth]{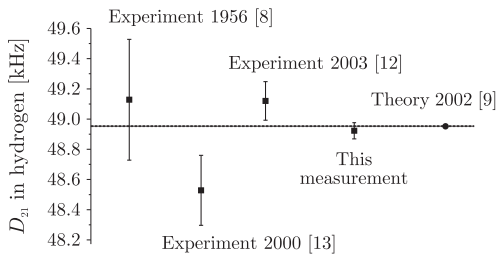}
\caption{Experimental and theoretical values for the $D_{21} = 8
f_{\rm HFS}({2S}) - f_{\rm HFS}({1S})$ difference in atomic
hydrogen.}\label{fig4}
\end{center}
\end{figure}

\paragraph{The AC Stark shift.} The differential AC Stark shift of the
$2S$ hyperfine components is  negligible ($\sim 1\,\mu$Hz/mW
\cite{Kolachevsky04}). Though the measured difference in {\it
singlet} and {\it triplet} excitation powers is taken into account
for each individual spectrum, the uncertainty of correction itself
may cause an error in $f_\textrm{HFS}(2S)$. Comparison of the data
evaluated with/without correction shows a systematic difference
from 2\,Hz to 5\,Hz depending on the delay $\tau$. Assuming an
error in the AC Stark shift evaluation of 30\,\% (including the
error in the power measurement), the contribution to the final
uncertainty budget is 1.3\,Hz.

\paragraph{The DC Stark shift.} We have no possibility to measure
stray electric fields in our apparatus and evaluate its
contribution of $-1(1)$\,Hz as in \cite{Kolachevsky04}. The
influence of the quench fields 1,\,2 is tested with the help of
 simulations \cite{Kolachevsky06}, an effect on the
sub-hertz level is expected.

\paragraph{Magnetic fields.} The sensitivity of $f_\textrm{HFS}(2S)$ to an external magnetic
field $B$ equals $+9\,600\,B^2\,\textrm{Hz/G}^2$.
In 5 sets of measurements the compensation field direction was
inverted which increased the field in the less shielded zone from
30\,mG to 300\,mG (Fig.\,\ref{fig2}). The corresponding
$f_\textrm{HFS}(2S)$ differs from the value measured with proper
orientation by $-3(12)$\,Hz at $\tau=810\,\mu$s. We estimate the
uncertainty resulting from magnetic fields as 0.5\,Hz.

Summarizing  the uncertainties (Table\,\ref{t1}) we get the final
result $f_\textrm{HFS}(2S)=177\,556\,834.3(6.7)$\,Hz. The
corresponding value $D_{21}=48\,923(54)$\,Hz is in good agreement
with the theoretical prediction (Fig.\,\ref{fig4}) which is the
first result after \cite{Kusch56} consistent with theory within
$1\,\sigma$. This measurement is part of the long-term project on
precision spectroscopy of the $1S$\,-\,$2S$ transition in H
\cite{Fischer04} and has an impact on its centroid frequency
 used for the Rydberg constant derivation \cite{Biraben08}.

\begin{table}[t!]
\caption{Uncertainty budget for the new $2S$ HFS frequency
measurement in atomic hydrogen. }
\begin{tabular}{l c c }

  \hline
  \hline
    & Frequency [Hz] & Uncertainty [Hz] \\

  \hline
 Averaged interval frequency & 177 556 835.3 & 6.2 \\
 Line shape/temperature & 0 & 2 \\
 DC Stark shift & -1 & 1 \\
 AC Stark shift & 0 & 1.3 \\
 Magnetic fields & 0 & 0.5 \\
 \hline
 Final result & 177 556 834.3 & 6.7 \\
\hline \hline
\end{tabular}
\label{t1}
\end{table}

The work is supported in part by RSSF, MD-887.2008.2, DFG GZ 436
RUS 113/769/0-3 and RFFI 08-02-91969. J.A. is supported by EU
Marie Curie fellowship. The authors are grateful to Th.\,Udem and
G.\,Gabrielse for stimulating discussions.



\end{document}